\documentstyle[aps,12pt,epsf]{revtex}
\def\la{\mathrel{\mathpalette\fun <}}
\def\ga{\mathrel{\mathpalette\fun >}}
\def\fun#1#2{\lower3.6pt\vbox{\baselineskip0pt\lineskip.9pt
  \ialign{$\mathsurround=0pt#1\hfil##\hfil$\crcr#2\crcr\sim\crcr}}}

\def\nualpha{{\nu_\alpha}}

\def\nubeta{{\nu_\beta}}

\begin{document}

\draft
\title{Neutrino-Mixing-Generated Lepton Asymmetry
and the Primordial $^4$He Abundance}

\author{Xiangdong~Shi, George~M.~Fuller and Kevork~Abazajian}
\address{Department of Physics, University of California,
San Diego, La Jolla, California 92093-0319}

\date{January 23, 1999}

\maketitle

\begin{abstract}
It has been proposed that an asymmetry in the
electron neutrino sector may be generated by
resonant active-sterile neutrino transformations
during Big Bang Nucleosynthesis (BBN). We calculate
the change in the primordial $^4$He yield $Y$ resulting
from this asymmetry, taking into account both the time 
evolution of the $\nu_e$ and $\bar\nu_e$ distribution
function and the spectral distortions in these.
We calculate this change in two schemes: (1) a lepton
asymmetry directly generated by $\nu_e$ mixing with a
lighter right-handed sterile neutrino $\nu_s$; and 
(2) a lepton asymmetry generated by a 
$\nu_\tau\leftrightarrow\nu_s$ or
$\nu_\mu\leftrightarrow\nu_s$ transformation which
is subsequently partially converted to an asymmetry
in the $\nu_e\bar\nu_e$ sector by a matter-enhanced 
active-active neutrino transformation. In the first
scheme, we find that the percentage change in $Y$ is
between $-1\%$ and $9\%$ (with the sign depending on
the sign of the asymmetry), bounded by the Majorana 
mass limit $m_{\nu_e}\la 1$ eV. In the second scheme,
the maximal percentage reduction in $Y$ is $2\%$, if 
the lepton number asymmetry in neutrinos is positive;
Otherwise, the percentage increase in $Y$ is $\la 5\%$
for $m^2_{\nu_\mu,\nu_\tau}-m^2_{\nu_s}\la 10^4$~eV.
We conclude that the change in the primordial $^4$He
yield induced by a neutrino-mixing-generated lepton 
number asymmetry can be substantial in the upward 
direction, but limited in the downward direction.

\end{abstract}
\bigskip


\newpage
\section{Introduction}
It has been known that a resonant active-sterile
neutrino transformation during the Big Bang
Nucleosynthesis (BBN) epoch can generate
lepton number asymmetries in the active 
neutrino sectors \cite{Foot1,Shi2,Foot2,Shi3}.
The generated lepton number asymmetry $L_{\nu_\alpha}$
(with $\nu_\alpha$ being any of the three active neutrino
species) has an order of magnitude
\begin{equation}
L_{\nu_\alpha}\equiv {n_{\nu_\alpha}-n_{\bar\nu_\alpha}\over n_\gamma}
\sim \pm{\vert\delta m^2/{\rm eV}^2\vert\over 10\,(T/{\rm MeV})^4}
\quad{\rm for\ }\vert L_{\nu_\alpha}\vert\ll 0.1.
\label{asymptotic}
\end{equation}
In this equation, $n$ is the particle proper number density,
$\delta m^2\equiv m^2_{\nu_s}-m^2_{\nu_\alpha}$,
and $T$ is the temperature of the universe.

The asymmetry may have an appreciable impact on the
the primordial $^4$He abundance $Y$ if it is in the
$\nu_e$ sector and if its magnitude at the weak freeze-out
temperature $T\sim 1$ MeV is $\vert L_{\nu_e}\vert\ga 0.01$.
The resulting change of the primordial $^4$He abundance, 
$\Delta Y$, however, is not easy to estimate.  Not only
is the generated asymmetry a function of time, but also
the $\nu_e$ or $\bar\nu_e$ energy spectrum is distorted
by mixing (in a time-dependent fashion as well).
Therefore, previous attempts \cite{Foot3} to estimate
$\Delta Y$ in this lepton number generation scenario,
employing BBN calculations based on a constant asymmetry
and a thermal neutrino spectrum for electron-type 
neutrinos, is overly simplistic and may yield 
inaccurate results.

In this paper, we discuss in detail the time 
evolution of $L_{\nu_e}$ and the distortion
in the $\nu_e$ or $\bar\nu_e$ spectrum in this 
neutrino-mixing-driven lepton number asymmetry
generating scenario. 
We then calculate $\Delta Y$ taking into
account these time-dependent and energy-dependent
effects by modifying the standard BBN code
accordingly. We consider two schemes:
a direct one and an indirect one. 

The direct scheme involves a direct resonant
$\nu_e\leftrightarrow\nu_s$ transformation
which generates $L_{\nu_e}$ \cite{Foot1,Shi2}.
The indirect one first has a lepton asymmetry 
generated via a resonant $\nu_\mu\leftrightarrow\nu_s$
or $\nu_\tau\leftrightarrow\nu_s$
transformation and then has the asymmetry partially
transferred into $L_{\nu_e}$
by an active-active $\nu_\mu\leftrightarrow\nu_e$ or
$\nu_\tau\leftrightarrow\nu_e$ transformation \cite{Foot3}.
In both cases, Eq.~(\ref{asymptotic}) indicates
that the active-sterile channel requires 
$m^2_{\nu_\alpha}-m^2_{\nu_s}\ga 0.1$ eV$^2$
to generate $\vert L_{\nu_\alpha}\vert\ga 0.01$
at a temperature $T\sim 1$ MeV. This implies
$m_{\nu_\alpha}\ga 0.3$ eV.  In addition,
the effective mixing angles associated with
these neutrino oscillation channels have to
be large enough to generate lepton number 
asymmetries efficiently, but not so large 
as to produce too many $\nu_s$'s before or
during the onset of the process. The excess
$\nu_s$'s may not only suppress the lepton
number generation, but also yield a $^4$He
mass fraction that is too large \cite{Shi2,Shi1}.
These conditions can be quantified as:
\begin{equation}
\vert\delta m^2/{\rm eV}^2\vert^{1/6}\sin^22\theta\ga 10^{-11},
\label{lowlim}
\end{equation}
\begin{equation}
\vert\delta m^2/{\rm eV}^2\vert\sin^42\theta\la 10^{-9}\,(10^{-7})\quad
{\rm for}\ \nu_\alpha=\nu_e\,(\nu_\mu,\,\nu_\tau).
\label{BBNlim}
\end{equation}
where $\theta$ is the vacuum mixing angle.

The change in the predicted primordial $^4$He abundance
can go either way, depending on whether $L_{\nu_e}$ is positive
(decreasing $Y$) or negative (increasing $Y$). Our detailed 
BBN calculations show that in the direct scheme, we have
$-0.002\leq\Delta Y\leq 0.022$, bounded by the $\nu_e$
Majorana mass limit $m_{\nu_e}\la 1$ eV. This possible change
in $Y$ is significant compared to the uncertainty involved in 
the $Y$ measurements.
(The measured primordial $^4$He abundance from Olive {\sl et al.}
is $Y=0.234\,\pm 0.002\,(stat.)\,\pm 0.005\,(syst.)$\cite{Olive},
while another group claims $Y=0.244\pm 0.002$ \cite{Izotov}.) 
In fact, an increase in $Y$ of magnitude $\ga 0.01$ due to a
large negative $L_{\nu_e}$ would already be inconsistent
with observations. In the indirect scheme, we find that the maximal
possible reduction in $Y$ is 0.005. The expected increase of $Y$
goes up with $m^2_{\nu_\mu,\nu_\tau}-m^2_{\nu_s}$, and can be as
high as 0.013 for $m^2_{\nu_\mu,\nu_\tau}-m^2_{\nu_s}\sim 10^4$ eV$^2$.
In both schemes, $\Delta Y$ is rather limited in the negative
direction. The possible reduction in $Y$, resulting from a positive
$L_{\nu_e}$, may to some degree narrow the gap between the lower $Y$
measurement \cite{Olive} and the standard BBN prediction ($Y=0.246\pm 0.001$
\cite{Turner} assuming a primordial deuterium abundance 
D/H$\approx 3.4\pm 0.3\times 10^{-5}$\cite{Tytler}).
We note, however, that the maximal reduction of
$\Delta Y\approx -0.005$ is achieved only in the indirect
scheme when $m^2_{\nu_\mu,\nu_\tau}-m^2_{\nu_s}\sim 100$
to 300 eV$^2$, implying an unstable $\nu_\mu$
or $\nu_\tau$ with $m_{\nu_\mu,\nu_\tau}\ga 15$~eV\cite{Shi3}.

\section{Generation of Lepton Asymmetry by Resonant 
Active-Sterile Neutrino Transformation}

The formalism of active-sterile neutrino transformation and
associated amplification of lepton asymmetry when $\delta m^2
\equiv m_{\nu_s}^2-m_{\nu_\alpha}^2<0$ has been discussed
extensively elsewhere \cite{Foot1,Shi2,Foot2,Shi1,Enqvist1}.
Here we summarize
the main conclusions of these papers, and then concentrate our
discussion on the final stage of the amplification process, 
when $T$ approaches the weak freeze-out temperature $\sim$1 MeV.

For $\nu_\alpha\leftrightarrow\nu_s$ transformation
with $\delta m^2<0$, a resonance occurs at
\begin{equation}
T_{\rm res}\approx T_0\,\left({E\over T}\right)^{-1/3}
\left\vert{\delta m^2\cos 2\theta\over 1{\rm eV}^2}\right\vert^{1/6},
\label{Tres}
\end{equation}
where $T_0\approx 19(22)$ MeV for $\alpha=e\,(\mu,\tau)$, and $E/T$
is the neutrino energy normalized by the ambient temperature.
Below $T_{\rm res}$, lepton asymmetry may be amplified to 
asymptotically approach one of the two values in eq.~(\ref{asymptotic}).
Before this asymptotic value is reached, however, there is a brief
\lq\lq chaotic\rq\rq phase in which $L_{\nu_\alpha}$ oscillates
around zero \cite{Shi2}. As a result, the sign of $L_{\nu_\alpha}$
that emerges from the chaotic phase is unpredictable. We should
point out that the detailed numerical evolution of the generated
lepton number remains controversial. For example, whether or not
the evolution of the lepton number represents true chaos is not
precisely known. However, our BBN arguments simply reply on the
sensitivity of the generated lepton number to the neutrino oscillation
parameters. This sensitivity leads to a causality consideration on
the sign of the generated lepton number. A discussion of the
causality consideration can be found in Ref. \cite{ShiPRL}.

It is not surprising that at these two asymptotic $L_{\nu_\alpha}$
values, either $\nu_\alpha\leftrightarrow\nu_s$ (if $L_{\nu_\alpha}<0$)
or $\bar\nu_\alpha\leftrightarrow\bar\nu_s$ (if $L_{\nu_\alpha}>0$)
undergoes resonant transition due to matter effects. The system
maintains the growth of $L_{\nu_\alpha}$ by converting one of
$\nu_\alpha$/$\bar\nu_\alpha$ resonantly but suppressing the
transformation of the other. Of course, not all $\nu_\alpha$ or 
$\bar\nu_\alpha$ undergo the resonant transformation, because 
neutrinos in the early universe have an energy distribution
and the resonance condition is energy dependent.
When $L_{\nu_\alpha}$ is small, even the resonant conversion
of a small fraction of either $\nu_\alpha$ or $\bar\nu_\alpha$
with energy $E_{\rm res}$ will be enough to maintain the growth
of $L_{\nu_\alpha}$.  To quantify the above arguments, we note
that the effective potential {\bf V}$=(V_x,\,V_y,\,V_z)$ of
the $\nu_\alpha\leftrightarrow\nu_s$ transformation channel is
\begin{equation}
V_x=-{\delta m^2\over 2E}\sin 2\theta,\quad
V_y=0,\quad
V_z=-{\delta m^2\over 2E}\cos 2\theta+V_\alpha^L+V_\alpha^T.
\label{Vcomponent}
\end{equation}
The contribution from matter-antimatter 
asymmetries (matter effect) is \cite{Raffelt}
\begin{equation}
V_\alpha^L\approx \pm 0.35 G_FT^3\Bigl[L_0
                  +2L_{\nu_\alpha} +\sum_
                  {\nubeta\neq\nualpha}L_{\nu_\beta}\Bigr],
\label{VL}
\end{equation}
where $G_F$ is the Fermi constant, and $L_0\sim 10^{-9}$
represents the contributions from the baryonic asymmetry
as well as the asymmetry in electron-positions.
The \lq\lq$+$\rq\rq\ sign is for the neutrino oscillation
channel, and the \lq\lq$-$\rq\rq\ sign is for the anti-neutrino
oscillation channel.  The contribution from the thermal neutrino
background is $V_\alpha^T$, whose value is \cite{Raffelt}
\begin{equation}
V_\alpha^T \approx -A{n_{\nu_\alpha}+n_{\bar\nu_\alpha}\over
n_\gamma}G_F^2ET^4,
\label{VT}
\end{equation}
where $A\approx 110 (30)$ for $\alpha=e\,(\mu\ {\rm  or}\ \tau)$.
The effective matter mixing angle at temperature $T$ is
\begin{equation}
\sin 2\theta_{\rm eff}={V_x\over (V_x^2+V_z^2)^{1/2}},
\label{angle}
\end{equation}
which reduces to vacuum mixing when $V_\alpha^L$ and $V_\alpha^T$ are zero.

Several physical processes with different times scales 
are involved in the resonant 
$\nu_\alpha\leftrightarrow\nu_s$ transformation process:
(1) the local neutrino oscillation rate $\vert{\bf V}\vert$;
(2) the weak interaction rate $\Gamma\sim 4G_F^2T^5$;
(3) the Hubble expansion rate $H=-\dot T/ T=5.5T^2/m_{\rm pl}$
where $m_{\rm pl}\approx 1.22\times 10^{28}$ eV is the Planck mass;
(4) the rate of change of $\vert{\bf V}\vert$, $\vert\dot{\bf V}\vert
/\vert{\bf V}\vert$, caused by the change of lepton asymmetry and the
Hubble expansion. If any one of the rates is much larger than the others,
we may consider all the other processes as perturbations, which simplifies
the picture greatly.  For example, if $\vert{\bf V}\vert$ dominates we may
consider the system as an ordinary neutrino transformation system with an 
effective mixing angle as in Eq.~(\ref{angle}), and with all physical 
variables changing adiabatically. If the weak interaction dominates, 
each weak scattering acts as a \lq\lq measurement\rq\rq\ to the mixing
system, effectively reducing the mixture to flavor eigenstates 
$\nu_\alpha$ and $\nu_s$.  The neutrino transformation is hence
suppressed, with a reduced $\sin 2\theta_{\rm eff}=V_x/(\Gamma/2)$.
If the Hubble expansion dominates, the other 
processes are essentially \lq\lq frozen out\rq\rq . This is
the case at $T\la 1$ MeV when the two-body weak interaction
freezes out and the neutron-to-proton ratio becomes fixed 
(other than from the free neutron decay process). If
$\vert\dot{\bf V}\vert$ dominates over $\vert{\bf V}\vert$,
neutrino amplitude evolution becomes non-adiabatic.

The ratios of the first three rates are:
\begin{equation}
\left\vert{V_x\over \Gamma}\right\vert
\approx 5\times 10^8\,\left({\vert\delta m^2\vert\over
{\rm eV}^2}\right)\,\left({T\over{\rm MeV}}\right)^{-6}\,\sin 2\theta;
\label{ratio1}
\end{equation}
\begin{equation}
\left\vert{V_x\over H}\right\vert
\approx 10^9\,\left({\vert\delta m^2\vert\over
{\rm eV}^2}\right)\,\left({T\over{\rm MeV}}\right)^{-3}\,\sin 2\theta;
\label{ratio2}
\end{equation}
\begin{equation}
{\Gamma\over H}\approx \left({T\over{\rm MeV}}\right)^3.
\label{ratio3}
\end{equation}
In these ratios, an average $E=3.15T$ is assumed. Since we
are only concerned with $\vert\delta m^2\vert\ga 0.1$ eV$^2$,
as long as $\sin 2\theta\ga 10^{-6}$ (the minimal mixing required
to amplify $L_{\nu_\alpha}$, see Eq.~({\ref{lowlim})), the neutrino
transformation rate easily dominates over the Hubble expansion and
the weak interaction at $T\sim 1$ MeV. 

We always have ${\dot V_x}=HV_x$.
At $T\la T_{\rm res}/2$ and away from resonances so that
we can assume $V_\alpha^T\ll\vert\delta m^2\vert/2E
\sim V_\alpha^L\sim V_z$, we have $\vert{\dot V_z}\vert
\sim \vert (H-{\dot L_{\nu_\alpha}}/L_{\nu_\alpha})V_z\vert
\sim \vert HV_z\vert$. (Note that $L_{\nu_\alpha}$ is limited
to the $T^{-4}$ growth in Eq.~(\ref{asymptotic}).)
Therefore at $T\la T_{\rm res}/2$,
the active-sterile neutrino transformation channels 
in our problem can be treated as ordinary oscillation
channels with adiabatically varying mixing parameters
except possibly at the resonances. We will discuss
the question of adiabaticity at resonances later.

At $T\la T_{\rm res}/2$ when we can neglect
$V_\alpha^T$, the resonance condition
$V_z=-\delta m^2/2E\cos 2\theta + V_\alpha^L=0$ gives the
asymptotic values of $L_{\nu_\alpha}$ in Eq.~(\ref{asymptotic}).
Note that for $L_{\nu_\alpha}<0$ ($L_{\nu_\alpha}>0$) the
$\nu_\alpha$ ($\bar\nu_\alpha$) transformation channel is
matter-enhanced. The fraction $F$ of resonantly converted
$\nu_\alpha$ ($\bar\nu_\alpha$) in the total $\nu_\alpha$
($\bar\nu_\alpha$) distribution is
\begin{equation}
F\sim 2\left\vert V_x{{\rm d}\epsilon\over
{\rm d}V_z}\right\vert_{\epsilon_{\rm res}}
f(\epsilon_{\rm res})\approx
2\epsilon_{\rm res}\sin 2\theta f(\epsilon_{\rm res})
\end{equation}
where $\epsilon\equiv E/T$, 
$f(\epsilon)\approx 2\epsilon^2/3\zeta(3)(1+e^{\epsilon})$
and $\epsilon_{\rm res}$ satisfies $V_z(\epsilon_{\rm res})=0$.
The resonance $\epsilon_{\rm res}$ remains stationary
(or varies very slowly) and $L_{\nu_\alpha}\propto T^{-4}$ if
\begin{equation}
\vert L_{\nu_\alpha}\vert \la {3\over 8}F\sim
{3\over 4}\epsilon_{\rm res}\sin 2\theta f(\epsilon_{\rm res})
\sim 0.1\sin 2\theta.
\label{stationary}
\end{equation}
In general, the stationary $\epsilon_{\rm res}$
has a value $\sim {\cal O}(0.1)$\cite{Foot2,Shi3}.
When $\vert L_{\nu_\alpha}\vert >0.1\sin 2\theta$, 
$\vert L_{\nu_\alpha}\vert$ cannot keep up with 
the $T^{-4}$ growth. Consequently $\epsilon_{\rm
res}$ must increase as $\epsilon_{\rm res}\propto
T^{-4}\vert L_{\nu_\alpha}\vert^{-1}$ at low
temperatures when $V^T$ can be safely neglected.

The resonance is adiabatic when the resonance region is
characterized by slowly changing $\epsilon_{\rm res}$.
As the resonance sweeps through the neutrino energy
spectrum, a complete conversion of $\nu_\alpha$ to 
$\nu_s$ at the resonance is possible only if the 
following adiabatic conditions are met:
\begin{equation}
\left\vert {V_x^2\over \dot V_z^\prime}\right\vert\sim
\left\vert{V_x\sin 2\theta^\prime\over H}\right\vert
\sim  10^9\,\left({\vert\delta m^2\vert\over {\rm eV}^2}\right)\,
\left({T\over{\rm MeV}}\right)^{-3}\,\sin^2 2\theta\gg 1,
\label{Mixingnotoosmall}
\end{equation}
and
\begin{equation}
2\left\vert V_x{{\rm d}\epsilon
\over {\rm d}V_z}\right\vert_{\epsilon_{\rm res}}
f(\epsilon_{\rm res})\,
\left\vert{{\rm d}\epsilon_{\rm res}\over{\rm d} L_{\nu_\alpha}}\right\vert\ga
V_x\left\vert{{\rm d}\epsilon\over {\rm d}V_z}\right\vert_{\epsilon_{\rm res}}.
\end{equation}
The first condition simply implies that the timescale of completing
the resonance has to be much longer than the neutrino oscillation
period at resonance. This is satisfied if $\vert\delta m^2\vert
\sin^2 2\theta\ga 10^{-9}$ eV$^2$.  The second condition requires
that $\epsilon_{\rm res}$ move slowly through the spectrum.
This amounts to
\begin{equation}
L_{\nu_\alpha}\le {3\over 4}\epsilon_{\rm res}f(\epsilon_{\rm res}),
\end{equation}
which is satisfied for $\epsilon_{\rm res}$'s that cover the bulk
of the neutrino energy spectrum.  Adiabaticity at resonances is 
therefore a valid assumption for $\nu_\alpha\leftrightarrow\nu_s$
mixing with vacuum mixing angles which are not too small but
which are still within the BBN bound Eq.~(\ref{BBNlim}).

Given adiabaticity, the growth of $L_{\nu_\alpha}$ (assumed
to be positive for simplicity and no loss of generality) as
$\epsilon_{\rm res}$ sweeps through the neutrino energy
spectrum can be easily estimated by solving
\begin{equation}
L_{\nu_\alpha}(T)\approx {3\over 8}\int_0^{\epsilon_{\rm res}(T)}
\beta f(\epsilon){\rm d}\epsilon
\label{asymmetry1}
\end{equation}
and the resonance condition $V_z(\epsilon_{\rm res})=0$ which
at low temperatures is equivalent to
\begin{equation}
\epsilon_{\rm res}(T)\approx {\vert\delta m^2/{\rm eV}^2\vert
\over 16(T/{\rm MeV})^4L_{\nu_\alpha}}.
\label{asymmetry2}
\end{equation}
In Eq.~(\ref{asymmetry1}) $\beta$ takes account of the effect
of collisions that redistribute energy among neutrinos.

We can identify two extreme cases. When the collisions
are too inefficient to change the neutrino distribution
at $\epsilon >\epsilon_{\rm res}$ (such as when $T\la 1$
MeV), $\beta=1$.  In another limit, 
$\beta\approx 1-8L_{\nu_\alpha}/3$ which obtains when the 
collisions are highly efficient (such as when $T\ga 1$ MeV)
and neutrinos are always distributed thermally. 
Eqs.~(\ref{asymmetry1}) and (\ref{asymmetry2}) give a
solution in fair agreement with the results obtained
by solving Eq. (18) of Foot and Volkas \cite{Foot3}.
In Figure 1 we plot our results in terms of $L_{\nu_\alpha}$
vs. $m_{\nu_\alpha}^2-m_{\nu_s}^2$ at various temperatures.
From Figure 1 we can deduce a power law relation applicable
to $\vert L_{\nu_\alpha}\vert\la 0.1$,
\begin{equation}
\vert L_{\nu_\alpha}\vert\approx 0.05
\vert\delta m^2/{\rm eV}^2\vert^{2/3}\,(T/{\rm MeV})^{-8/3}.
\label{powerlaw}
\end{equation}
We note that this power law applies only in the stage when
the resonance sweeps through the neutrino energy spectrum.
When the resonance is stationary (Eq.~(\ref{stationary})),
the dependence is $\vert L_{\nu_\alpha}\vert\propto 
\vert\delta m^2\vert T^{-4}$ instead.

The asymmetry $L_{\nu_\alpha}$ is generated as the
resonance conversion region moves up through the 
neutrino energy spectrum.
This suggests a distortion of the $\nu_\alpha$ (or
$\bar\nu_\alpha$) energy spectrum (see also Ref. \cite{Chizhov}).
Indeed, when $L_{\nu_\alpha}\la 0.1$, most of the $L_{\nu_\alpha}$
is generated at the lowest temperatures (Eq.~[\ref{powerlaw}]),
when the neutrino scattering processes that tend to thermalize 
the neutrino spectrum are the most inefficient. The fact that
the resonant transformation of $\nu_\alpha$ to $\nu_s$
(or $\bar\nu_\alpha\rightarrow\bar\nu_s$) starts at lower
neutrino energies only further deepens the inefficiency of
neutrino re-thermalization, as neutrino interaction cross
sections scale roughly linearly with neutrino energies. In
Figure 2, we plot a semi-analytical calculation of the 
$\nu_\alpha$ spectrum at $T=1$ MeV for a 
$\nu_\alpha\rightarrow\nu_s$ resonant transformation
(which generates a negative $L_{\nu_\alpha}$) with
$m^2_{\nu_\alpha}-m^2_{\nu_s}=1$ eV$^2$. In the calculation,
the thermalization process is approximated as a relaxation
process (with a rate $\Gamma$) driving the system toward
a thermal distribution. As
a result of the inefficiency of this process, for cases
with $\vert\delta m^2\vert\la 1$ eV$^2$
(so that $L_{\nu_\alpha}\la 0.1$ at $T\sim 1$ MeV),
the $\nu_\alpha$ neutrino spectrum at and below its thermal
decoupling temperature $T\sim 1$ MeV can be well approximated
by a thermal spectrum with a low energy cut-off. The $\nu_\alpha$
deficit below the cut-off energy results in the $L_{\nu_\alpha}$
asymmetry. In the mean time, $\bar\nu_\alpha$
is not subject to resonant transformation. Its spectrum is
therefore not significantly changed, due to the inefficiency of
neutrino pair production. (The opposite is true if
$L_{\nu_\alpha}>0$: the $\bar\nu_\alpha$ distribution will have
its lower energy region truncated but the $\nu_\alpha$ distribution
will remain intact.)

\section{Direct Generation of Electron-Neutrino Asymmetry 
by Resonant {\rm $\nu_e\leftrightarrow\nu_s$} Transformations}

If $\alpha=e$ (the direct scheme to generate an asymmetry
in the $\nu_e\bar\nu_e$ sector), the $\nu_e$ or $\bar\nu_e$
spectral distortion will directly impact the neutron-to-proton
ratio at the weak freeze-out temperature, and hence the $^4$He
yield.  Figure 3 shows the changes in n$\leftrightarrow$p rates
due to the neutrino spectral distortion in the case
$m^2_{\nu_e}-m^2_{\nu_s}=1$ eV$^2$.
When $L_{\nu_e}>0$ (i.e., a deficit of low energy $\bar\nu_e$),
the major effect is an enhanced neutron decay rate at low
temperatures due to the reduced Pauli-blocking of $\bar\nu_e$.
For reaction p$+\bar\nu_e\rightarrow$n$+e^+$, the low energy
deficit in $\bar\nu_e$ is of little significance because only
$\bar\nu_e$ with $E>1.9$ MeV can participate in the reaction.
Conversely, its reverse reaction mostly generates $\bar\nu_e$
at the higher end of the energy spectrum. Its rate is therefore
insensitive to the spectral distortion at the lower end.
When $L_{\nu_e}>0$ (a deficit of low energy $\nu_e$),  the
rate for n$+\nu_e\rightarrow$p$+e^-$ is significantly reduced
while the reverse rate is slightly increased.
Figure 4 shows the resultant $\Delta Y$ from the spectral
distortion as a function of $m_{\nu_e}^2-m_{\nu_s}^2$ for
both $L_{\nu_e}>0$ and $L_{\nu_e}<0$. The disparity between
the two cases of opposite $L_{\nu_e}$ is transparent from
Figure 3: the change in n$\leftrightarrow$p rates is much
larger when $L_{\nu_e}<0$.

The $\nu_e$ Majorana mass limit (which is uncertain by a
factor of 2, ranging from $m_{\nu_e}\le 0.45$ eV to 
$m_{\nu_e}\la 1$ eV\cite{Acker,Guenter,Baudis}) implies
an upper limit $m_{\nu_e}^2-m_{\nu_s}^2\le 1$ eV$^2$.
\footnote{We point out that it might be possible to construct
mixings between left-handed electron neutrinos and {\sl left-handed}
sterile neutrinos.  In such a case the Majorana mass limit will not
apply but the standard weak interaction theory will then need to be
reconsidered.} Figure 4 shows that the maximally allowed reduction
in $Y$ is only $\approx 0.0021$, about 1\% of the standard
prediction. But the maximally allowed increase of $Y$ can be
as high as $\approx 0.022$, a 9\% effect. An increase this 
large in the predicted primordial $^4$He abundance would have
already been too large to accommodate observations\cite{Olive,Izotov}.

\section{Indirect Generation of Electron-Neutrino 
Asymmetry by Neutrino Transformations}

For convenience in the indirect scheme, we assume
that a $L_{\nu_\tau}$ is first generated by a
$\nu_\tau\leftrightarrow\nu_s$ transformation
process. This asymmetry may then be transferred to
$L_{\nu_e}$ by a resonant $\nu_\tau\leftrightarrow\nu_e$
oscillation \cite{Foot3}. (Ordinary oscillations without
resonance cannot transfer the asymmetry efficiently.)
In fact we are likely to have $\delta m^2_{(\tau s)}
\equiv m^2_{\nu_s}-m^2_{\nu_\tau}\approx\delta m^2_{(\tau e)}
\equiv m^2_{\nu_e}-m^2_{\nu_\tau}$ if
$m_{\nu_\tau}\gg m_{\nu_e},\,m_{\nu_s}$ (which will be the case
in order to have an appreciable impact on the primordial $^4$He
abundance). Such a neutrino mass spectrum is consistent with
a simultaneous solution of the solar neutrino problem 
\cite{Solar} and the atmospheric neutrino problem \cite{SuperK}.
For the moment, we will simply assume
$\delta m^2_{(\tau s)}\approx\delta m^2_{(\tau e)}$.

The matter asymmetry contributions to the effective potentials 
of the two neutrino transformation channels become
\begin{equation}
V_{(\tau s)}^L\approx \pm 0.35 G_FT^3(2L_{\nu_\tau}-L_{\nu_e}),\quad
V_{(\tau e)}^L\approx \pm 0.35 G_FT^3(L_{\nu_\tau}-L_{\nu_e}).
\label{twores}
\end{equation}
Apparently, the $\bar\nu_\tau\leftrightarrow\bar\nu_s$ and
$\bar\nu_\tau\leftrightarrow\bar\nu_e$ resonances when $L_{\nu_\tau}>0$
(or the $\nu_\tau\leftrightarrow\nu_s$ and $\nu_\tau\leftrightarrow\nu_e$
resonances when $L_{\nu_\tau}<0$) do not simultaneously share the same part
of the neutrino energy spectrum.

Guaranteed adiabaticity (i.e., satisfying Eq.~[\ref{Mixingnotoosmall}]),
the efficiency of resonant neutrino conversion is still determined by
whether or not the neutrino collision time scale dominates over the
timescale for a complete $\nu_\tau$ to $\nu_e$ (or $\bar\nu_\tau$ to
$\bar\nu_e$) transition at resonance (the resonance width). The
collision timescale is important because the two resonances in 
Eq.~(\ref{twores}) do not overlap, so for example, any deficit in
$\bar\nu_\tau$ caused by the $\bar\nu_\tau\leftrightarrow\bar\nu_s$
resonant transition relies on neutrino scattering to redistribute
neutrinos into the energy region where the
$\bar\nu_\tau\leftrightarrow\bar\nu_e$ resonance occurs.
In previous work \cite{Foot3} this redistribution has been assumed to
be instant. However, this is not a good approximation at $T\la 5$ MeV
as we will show below.

The ratio of the resonance timescale to the collision timescale is
\begin{equation}
\left\vert {V_x^{(\tau e)}\over \dot V_z^{(\tau e)}}\right\vert\,
\Gamma\sim {\Gamma\over H}\sin 2\theta_{(\tau e)}\sim
\left({T\over{\rm MeV}}\right)^3\sin 2\theta_{(\tau e)}.
\end{equation}
Calculations based on active-active neutrino
transformation in Type II supernova nucleosynthesis
suggest that $\sin^2 2\theta_{(\tau e)}\la 10^{-4}$
for $m^2_{\nu_\tau}-m^2_{\nu_e}\ga 1$ eV$^2$ \cite{Qian},
and the Bugey experiment constrains $\sin^2 2\theta_{(\tau e)}\le 0.04$
for $m^2_{\nu_\tau}-m^2_{\nu_e}\la 1$ eV$^2$ \cite{Bugey}.  
Therefore, the collision timescale dictates the growth of
$L_{\nu_e}$ at $T\la 5$ MeV for $\vert\delta m^2_{(\tau s)}
\vert\approx\vert\delta m^2_{(\tau e)}\vert\ga 1$ eV$^2$. 

Above 5 MeV, the collisions may be deemed instantaneous,
and the equations of ref. \cite{Foot3} become valid. But a
side effect of generating a large $L_{\nu_\tau}$ above 
$\sim 5$ MeV by the $\nu_\tau\leftrightarrow\nu_s$ mixing
(by having $m^2_{\nu_\tau}-m^2_{\nu_s}\ga 10^4$ eV$^2$)
is bringing sterile neutrinos into chemical equilibrium
\cite{Shi1}. As a result, even though a reduction as large
as $\Delta Y=-0.006$ may result from a positive $L_{\nu_e}$
alone in the indirect scheme \cite{Foot3}, the extra sterile
neutrinos produced will increase $Y$ by at least as much by 
bringing $\bar\nu_s$ into chemical equilibrium through the 
$\bar\nu_\tau\leftrightarrow\bar\nu_s$ resonant transformation.
The net effect is an increase of $Y$ instead of a reduction,
even if $L_{\nu_e}$ is positive.

We again model the transfer of $L_{\nu_\tau}$ to $L_{\nu_e}$ by
assuming that active neutrinos tend to their equilibrium distributions
with a rate $\Gamma$. Figure 5 shows the growth of $L_{\nu_\tau}$, 
$L_{\nu_e}$ and the accompanying increase in the total neutrino energy
density (including that of sterile neutrinos) in the units of that of one 
unperturbed active neutrino flavor, for one particular set of mixing
parameters. Figure 5 shows that $L_{\nu_\tau}$ indeed grows initially
according to Eq.~(\ref{powerlaw}), and tapers off at $L_{\nu_\tau}\approx
0.22$ when most of $\nu_\tau$ or $\bar\nu_\tau$ have undergone 
resonances. The transfer of $L_{\nu_\tau}$ to $L_{\nu_e}$ is 
efficient at $T\ga 2$ MeV, but freezes out below $\sim 2$ MeV.
It freezes out at a higher temperature than $L_{\nu_\tau}$
because the $\nu_\tau$-$\nu_e$ (or $\bar\nu_\tau$-$\bar\nu_e$)
resonance occurs at a higher energy than the $\nu_\tau$-$\nu_s$
(or $\bar\nu_\tau$-$\bar\nu_s$) resonance (Eq.~[\ref{twores}]).
The resonance therefore sweeps through the $\nu_e$ (or $\bar\nu_e$) 
spectrum faster.  The figure also shows that increase in the 
total neutrino energy density in this case
(about $2\%$, or $\Delta N_\nu\sim 0.07$) is moderate.

The spectra of $\nu_e$ and $\bar\nu_e$ in the indirect scheme
is only slightly distorted.  Figure 6 shows the modified
$\bar\nu_e$ spectrum when $L_{\nu_\tau},\,L_{\nu_e}>0$ for 
$m^2_{\nu_\tau}-m^2_{\nu_e}=m^2_{\nu_\tau}-m^2_{\nu_s}=
100$ eV$^2$, compared to an unperturbed active neutrino
spectrum with zero chemical potential. The distortion is 
small because the transfer of $L_{\nu_\tau}$ into $L_{\nu_e}$
occurs in the entire energy distribution (albeit at different
temperatures), unlike in the direct scheme when the generation
of $L_{\nu_e}$ occurs only in low energies. The distortion in
$\nu_e$ and $\bar\nu_e$ spectra can be well approximated by
an overall multiplication factor $1+\delta_{\pm}$. The net
asymmetry is therefore $L_{\nu_e}=3(\delta_{+}-\delta_{-})/8$,
and the percentage increase in the $\nu_e\bar\nu_e$ number 
density due to pair production is $(\delta_{+}+\delta_{-})/2$.  

By modifying the standard BBN code with the new $\nu_e$
and $\bar\nu_e$ spectra, and with the increased total 
neutrino energy density, we obtain their effects on $Y$
in Figure 7. At $m^2_{\nu_\tau}-m^2_{\nu_e}=m^2_{\nu_\tau}-m^2_{\nu_s}
\ll 100$ eV$^2$ the effect on $Y$ is dominated by
the asymmetry in the $\nu_e\bar\nu_e$ sector. But
as $m^2_{\nu_\tau}-m^2_{\nu_e}=m^2_{\nu_\tau}-m^2_{\nu_s}$
increases, the increased total neutrino energy density
gradually becomes significant. The increased energy density
causes $Y$ to increase, regardless of the sign of the
neutrino asymmetry.  As a result of these two factors,
a maximal reduction $\Delta Y\approx -0.005$ is achieved
in cases of positive lepton number asymmetries when 
$m^2_{\nu_\tau}-m^2_{\nu_e}=m^2_{\nu_\tau}-m^2_{\nu_s}
\sim 100$ to 300 eV$^2$. This mass-squared-difference,
however, implies that tau neutrinos are unstable, based
on cosmological structure formation considerations\cite{Shi3}.

Figure 7 is very different from the previous estimates
of Foot and Volkas \cite{Foot3}. For example, Foot and
Volkas have argued for a possible reduction
$\Delta Y\approx -0.006$ across the mixing
parameter range $10\la m^2_{\nu_\tau}-m^2_{\nu_e}
=m^2_{\nu_\tau}-m^2_{\nu_s}\la 3000$ eV$^2$. While
our result clearly shows a concave feature of $\Delta Y$
in this range, with a maximum at $\approx -0.005$.
Foot and Volkas' result also indicated that $\Delta Y$
is smaller in the positive direction (when $L_{\nu_\tau},\,L_{\nu_e}<0$)
than in the negative direction (when $L_{\nu_\tau},\,L_{\nu_e}>0$).
Our analysis indicates the opposite: when $L_{\nu_\tau},\,L_{\nu_e}<0$,
the changes in $Y$ due to the spectral asymmetry and the extra
neutrino energy add constructively; while $L_{\nu_\tau},\,L_{\nu_e}<0$,
these two effects add destructively. $\Delta Y$, therefore, is
larger in the positive direction than in the negative direction.

These differences, we believe, stem from our detailed consideration
of neutrino spectrum distortion and its time dependence.  These
factors are crucial to the neutron-to-proton freeze-out process,
and in turn the primordial $^4$He yield.  

\section{Summary}
We have calculated the spectral distortions for neutrinos and the
time dependence of the neutrino distribution function
during the lepton asymmetry generation via
resonant active-sterile neutrino transformation. We have included
these crucial effects in our BBN calculation assessing 
the effect on the primordial $^4$He abundance of the possible
lepton number asymmetry in the $\nu_e\bar\nu_e$ sector. We 
conclude that the possible increase in the primordial $^4$He
yield, as a result of a negative lepton number asymmetry, can
be substantial. The maximal increase can be as high as
$\sim 0.01$ to 0.02 (or 5 to 9\%) for mixing parameters 
that are consistent with neutrino mass constraints. The
possible decrease due to a positive lepton number asymmetry,
however, is limited to $\la 0.002$ (or $\la 1\%$) if the 
asymmetry is generated by a resonant
$\nu_e\leftrightarrow\nu_s$ mixing, or $\la 
0.005$ (or $\la 2\%$) if the asymmetry is generated by a 
three-family resonant mixing scheme. The magnitude of these
possible changes in the primordial $^4$He abundance induced
by the neutrino-mixing-generated lepton number asymmetry
is comparable to or greater than the uncertainty of current 
primordial $^4$He measurements. Therefore, the role of 
resonant active-sterile neutrino mixing in Big Bang 
Nucleosynthesis cannot be underestimated.

X.~S., G.~M.~F. and K.~A. are partially supported by NSF grant
PHY98-00980 at UCSD.

\newpage
\noindent{\bf Figure Captions:}

\noindent
Figure 1. The magnitude of lepton asymmetry
as a function of $\delta m^2$ at various temperatures. 
The bands denote the range of the asymmetry enclosed by the
two extreme cases: (1) collisions are completely inefficient 
(upper limits); (2) collisions are completely efficient (lower limits).
\bigskip

\noindent
Figure 2. The solid curve: the calculated $\nu_\alpha$ ($\alpha=e,\,
\mu,\,\tau$) distribution function. The dashed curve: an unperturbed
thermal neutrino distribution function with zero chemical potential.
\bigskip

\noindent
Figure 3. The change in n$\leftrightarrow$p rates due to $\bar\nu_e$
(if $L_{\nu_e}>0$) or $\nu_e$ (if $L_{\nu_e}<0$) spectral distortion
for $m^2_{\nu_e}-m^2_{\nu_s}=1$ eV$^2$.
\bigskip

\noindent
Figure 4. The impact on the primordial $^4$He abundance $Y$ if
an asymmetry in $\nu_e\bar\nu_e$ is generated by a resonant
$\nu_e\leftrightarrow\nu_s$ mixing in BBN. Baryon number density
to photon number density ratio is set to $\eta=5.1\times 10^{-10}$.
\bigskip

\noindent
Figure 5. The solid curves: $\vert L_{\nu_\tau}\vert$ and
$\vert L_{\nu_e}\vert$ as a function of temperature for
$m^2_{\nu_\tau}-m^2_{\nu_e}=m^2_{\nu_\tau}-m^2_{\nu_s}=100$
eV$^2$. The dashed curve: the increase in the total 
neutrino energy density as a function of temperature,
normalized by the energy density of one thermalized
active neutrino flavor with zero chemical potential.
\bigskip

\noindent
Figure 6. The solid curve: the calculated $\nu_e$ distribution
function in the indirect mixing scheme. The dashed curve: an
unperturbed thermal neutrino distribution function with zero
chemical potential. Inset: the ratio of the two distribution
functions vs. neutrino energy.
\bigskip

\noindent
Figure 7.  The impact on the primordial $^4$He abundance
$Y$ in the indirect neutrino mixing scheme, as a function
of $m^2_{\nu_\tau}-m^2_{\nu_e}=m^2_{\nu_\tau}-m^2_{\nu_s}$.
Baryon number density to photon number density ratio is set
to $\eta=5.1\times 10^{-10}$.
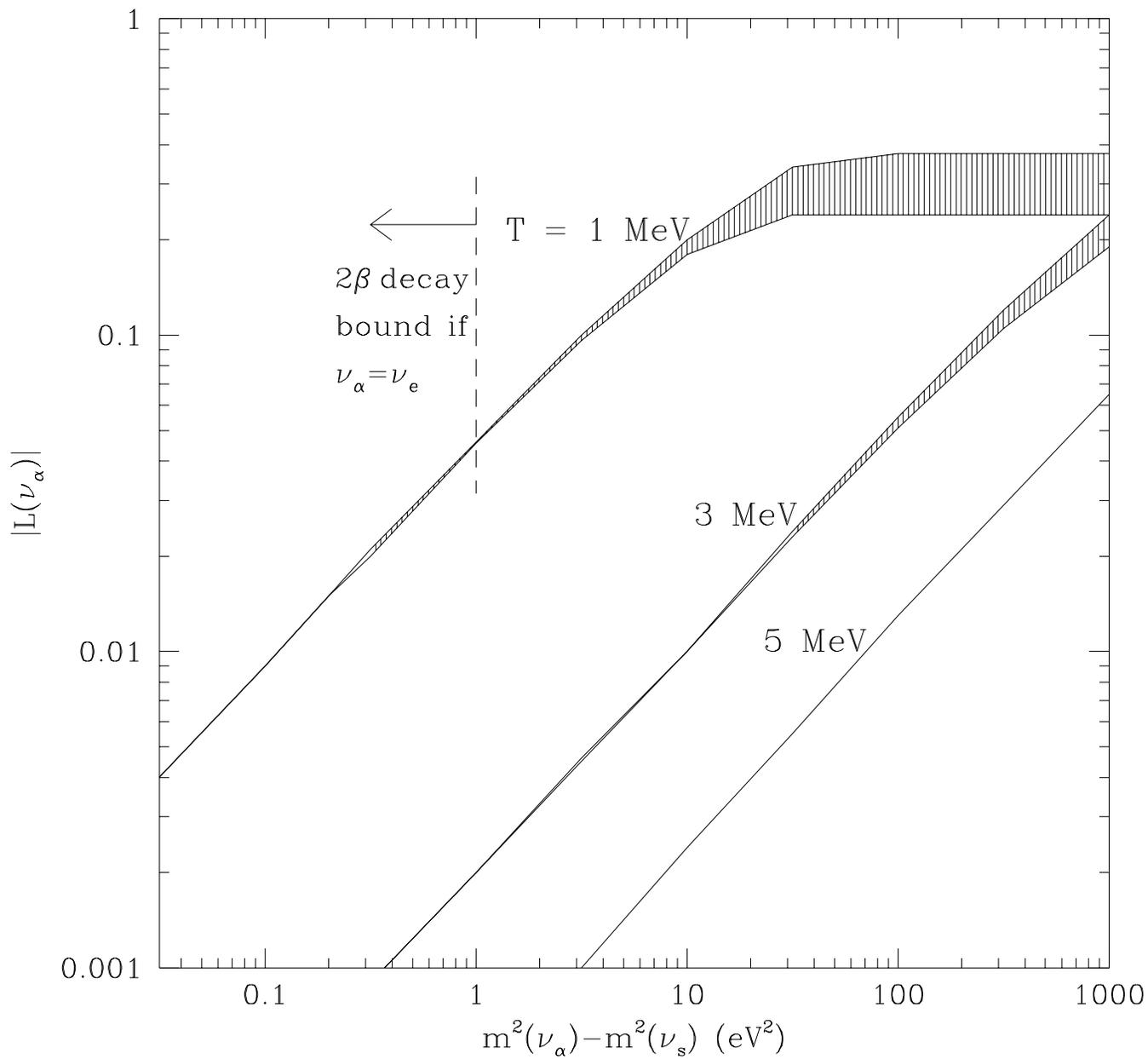
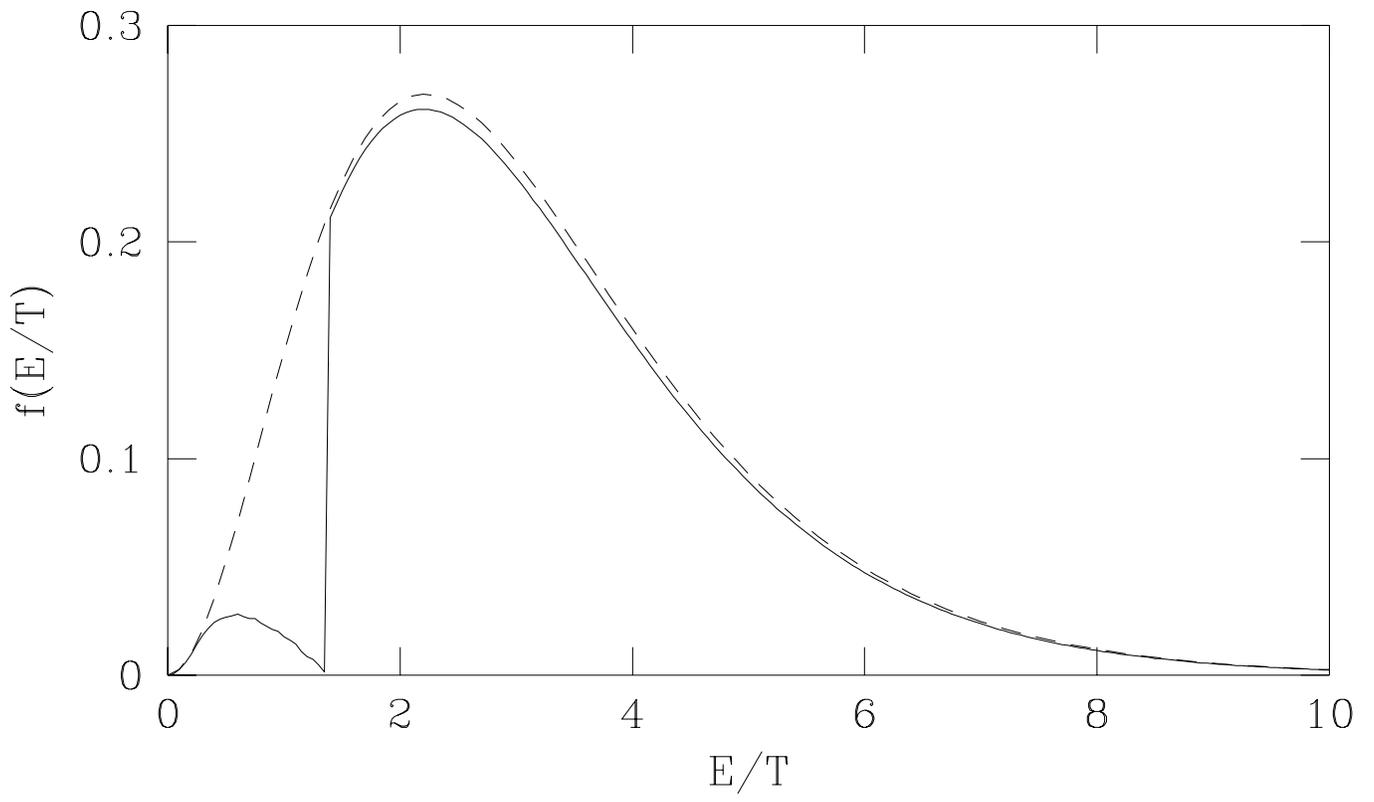
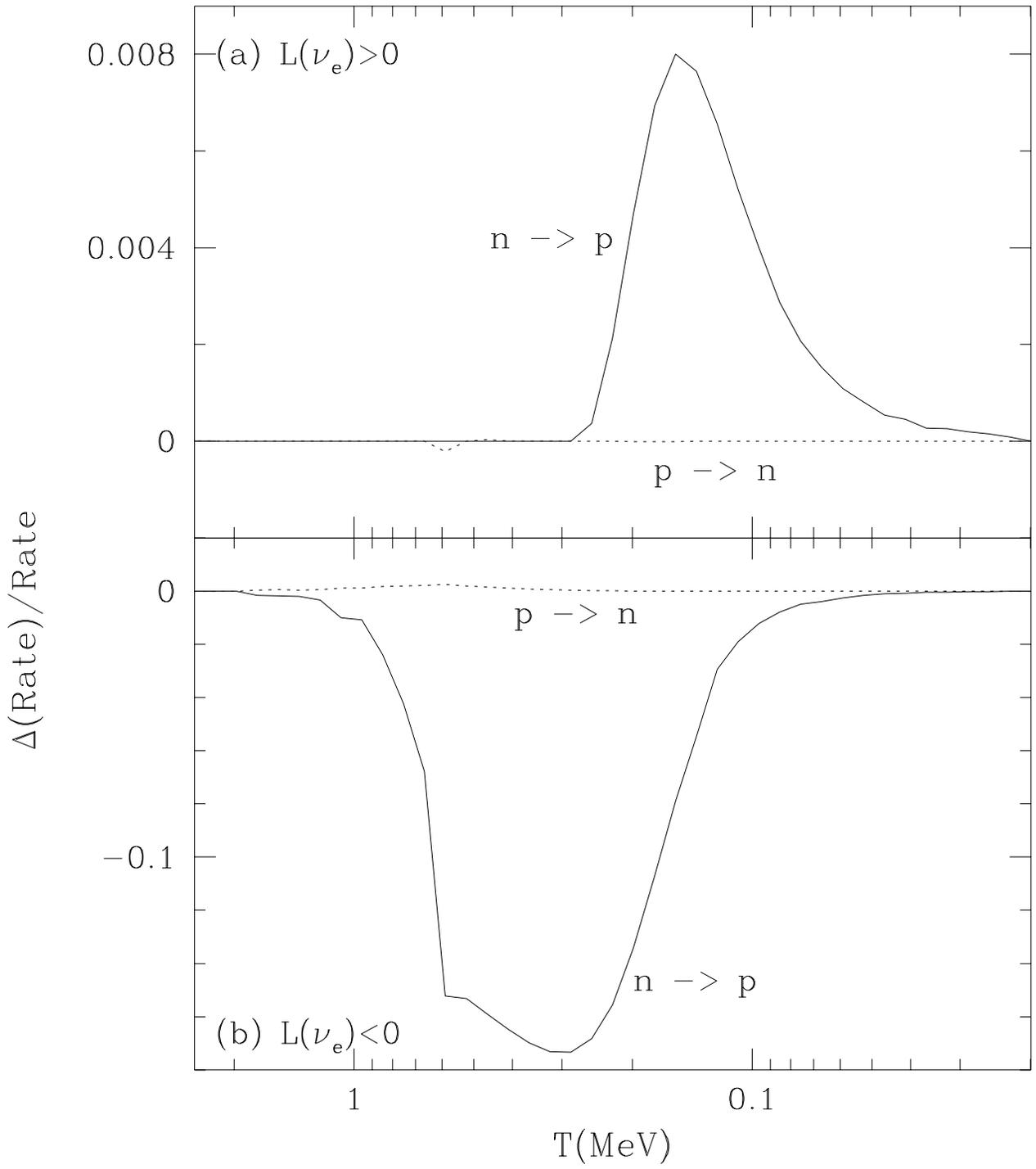
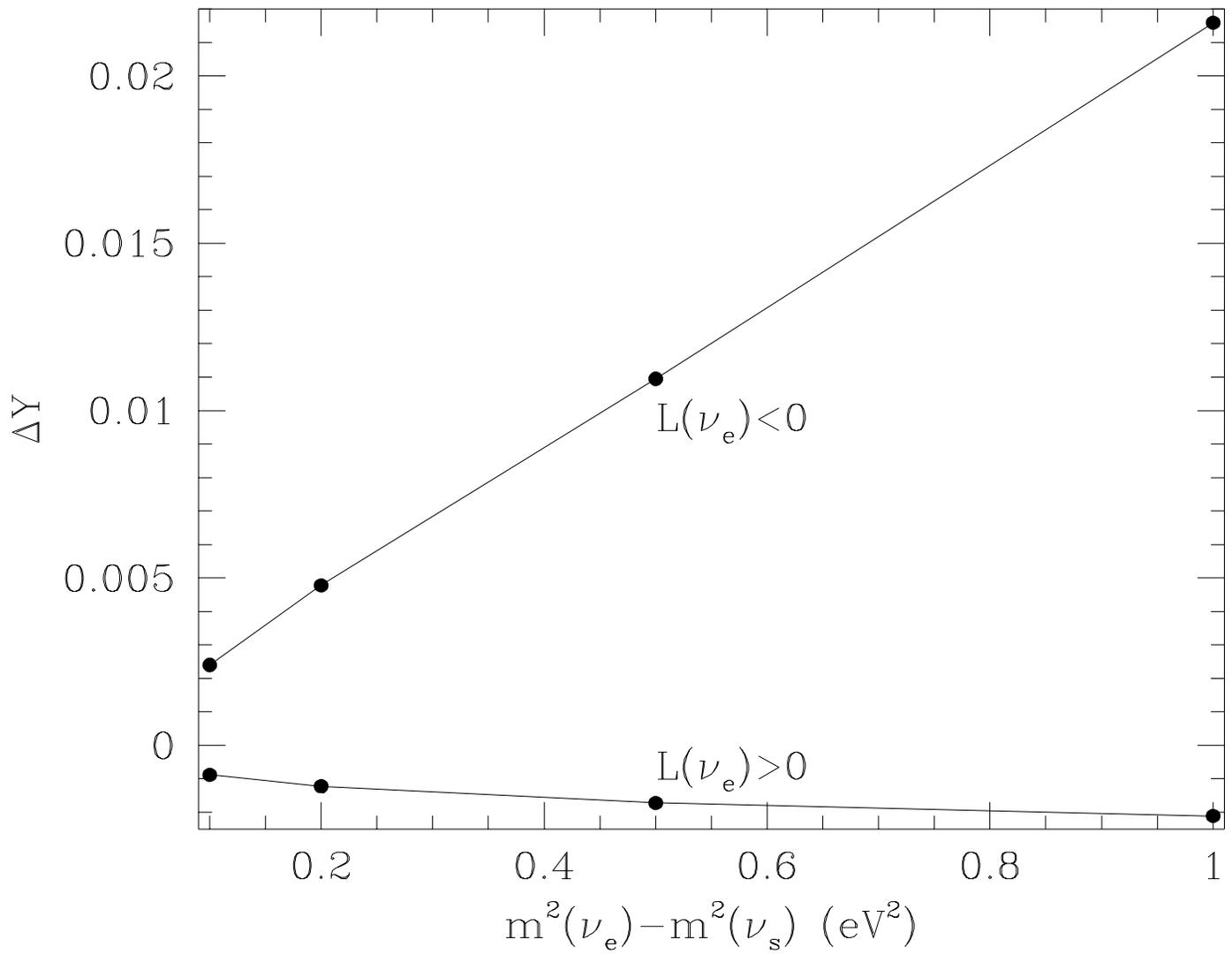
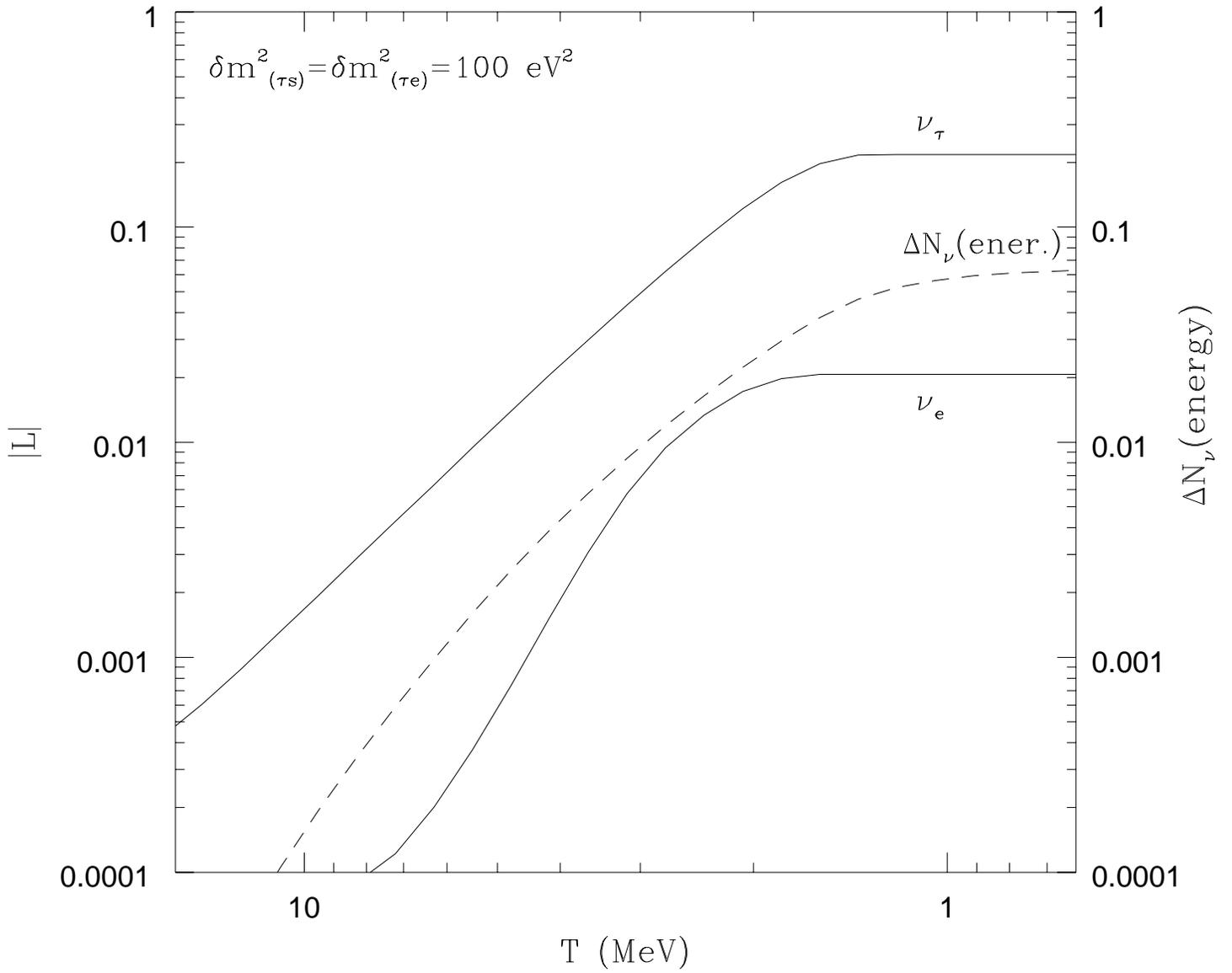
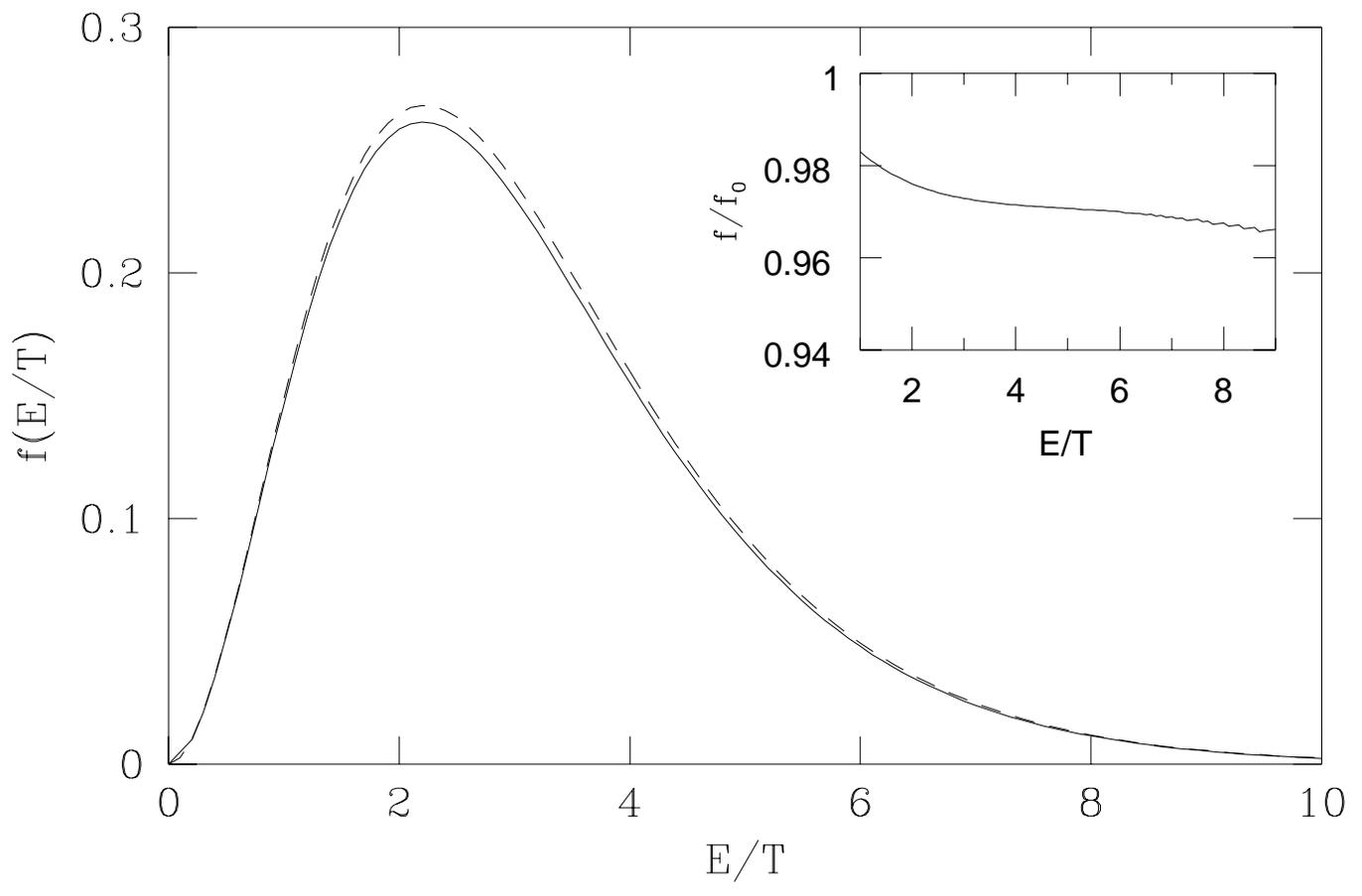
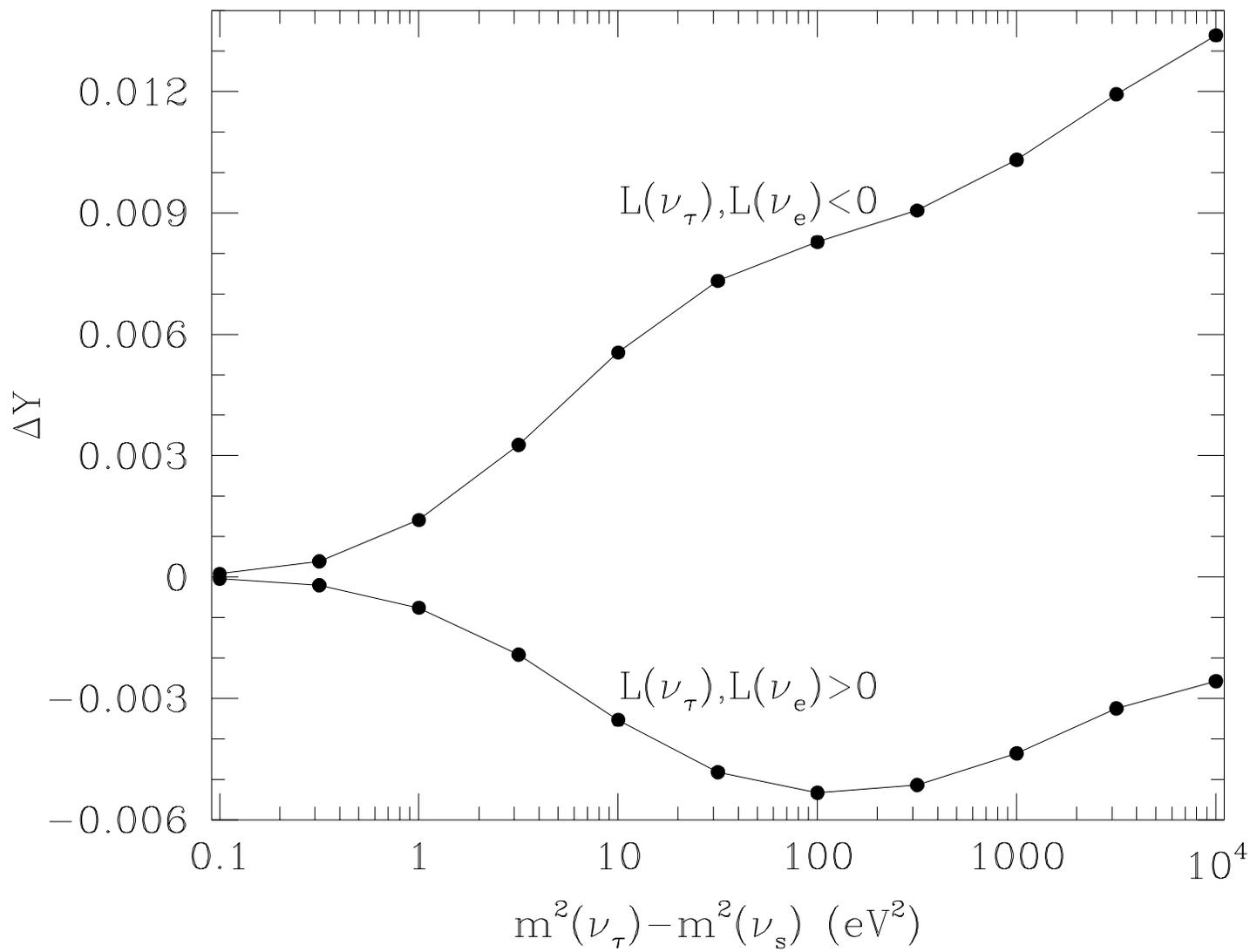
\end{document}